\documentclass[seceq,preprint]{jpsj2}

\def\runauthor{Kazumasa {\sc Miyake} and Jacques {\sc Flouquet}}

\title
{
Note on Anisotropy of Resistivity in Hidden Ordered State 
of URu$_2$Si$_2$
}

\recdate
{October 22, 2009}
\author
{Kazumasa {\sc Miyake} and Jacques {\sc Flouquet}$^{1}$}

\inst
{
Division of Materials Physics, Department of Physical Science, 
Graduated School of Engineering Science,\\ Osaka University, 
Toyonaka, Osaka 560-8531, Japan
\\
$^{1}$Commissariat \`a l' \'Energie Atomic, INAC, SPSMS, 
17 rue des Martyrs, 
38054 Grenoble, France
}

\abst
{
}

\kword
{URu$_2$Si$_2$, anisotropy of resistivity, hidden order, 
quadrupolar collective mode}

\begin{document}
\sloppy
\maketitle
\begin{figure}[h]
\begin{center}
\rotatebox{0}{\includegraphics[width=1.0\linewidth]{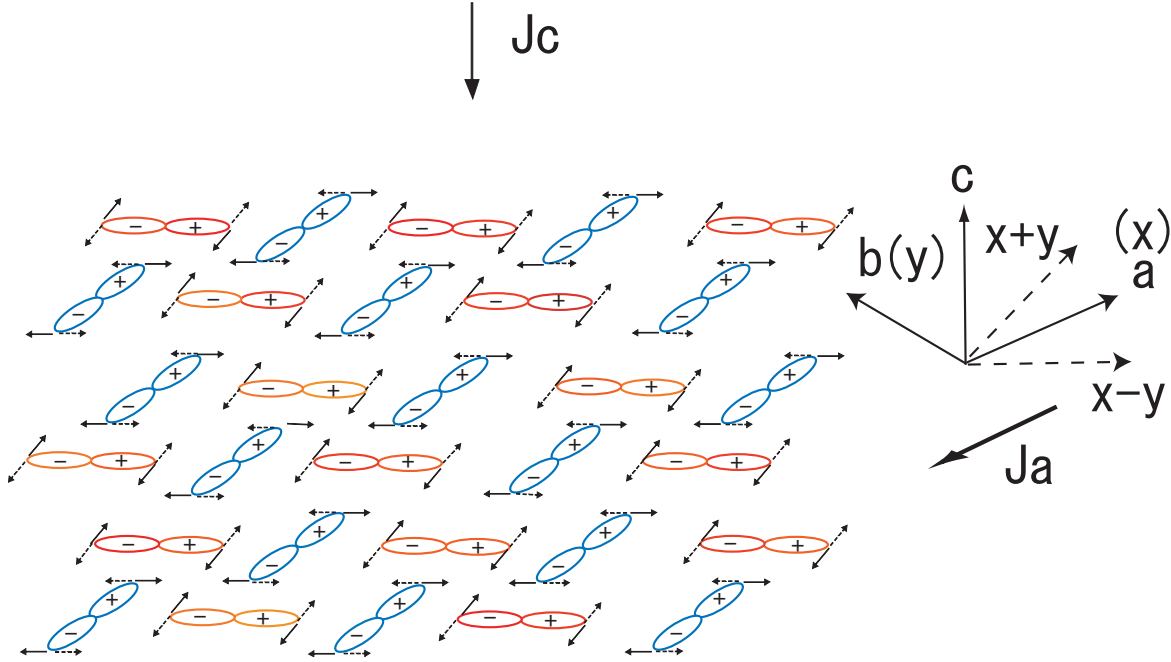}}
\caption{Coupling between AF-Q collective mode and current of electrons.  
The arrow of the solid line represents one of the directions of the rotation of induced local 
quadrupolar moments, $z(x+y)$ and $z(x-y)$, associated with the AF-Q collective 
mode in the long wavelength limit, and that of the dashed line the inverse direction 
of local quadrupolar moments.  Corresponding to the (0,0,1) order, red and blue 
dumbbells represent quadrupolar moments in $c=0$ and $c=1/2$ planes, respectively.  
}
\label{Fig:1}
\end{center}
\end{figure}

The original form of the so-called hidden order (HO) in URu$_2$Si$_2$ is still 
debated after its discovery a quarter century ago~\cite{Palstra}, 
while the properties of superconductivity~\cite{NMR,SpecificHeat,ThermalCond} 
and a large-moment antiferromagnetic phase (high-pressure phase)  
\cite{Sato,Amitsuka} have been clarified to some extent.  Recently, 
it has been observed by transport measurement that the resistivity 
along the $a$-axis, $\rho_{a}$, exhibits a non-Fermi liquid behavior, i.e., 
$\rho_{a}(T)\propto T$ in the low temperature limit, while that along the $c$-axis, 
$\rho_{c}$,  shows a typical Fermi liquid behavior, i.e., 
$\rho_{c}(T)\propto T^{2}$~\cite{Behnia}.  
This is a note that tries to explain this anisotropy of $T$ dependence 
on the assumption that the hidden order is the induced antiferro (AF)-quadrupolar 
(Q) order of $\Gamma_{5}$ symmetry, i.e., $Q_{z(x+y)}$ and $Q_{z(x-y)}$.  
The resulting induced quadrupolar charge distribution is of the O$_{xy}$ type, which is 
consistent with a state proposed recently by Harima and the present authors 
as an ordered state of the HO state~\cite{Harima}.  Note that 
$[z(x-y)]^{2} = z^{2}(x^{2}+y^{2}-2xy)$.  
A pattern of the wave function of quadrupolar ordering is shown schematically 
in Fig.\ \ref{Fig:1}.  
While quadrupolar ordering as an origin of the HO state has already been proposed 
by Santini and Amoretti\cite{QOrder1}, and Ohkawa and Shimizu\cite{QOrder2}, 
the state proposed by Harima {\it et al}. is consistent with almost all the hidden 
characteristics of the HO state~\cite{Harima} in contrast to the states proposed thus far.  
%The present type of ordering is also consistent with the fact that a resonance mode 
%at (1,0,0) observed by neutron scattering exists only in the hidden ordered 
%state~\cite{Flouquet}, as will be discussed elsewhere.  
%

The fundamental idea is rather simple.  The Nambu-Goldstone mode (collective mode  
in the AF-Q order) scatters electrons in different ways corresponding to the angle 
between the $c$-axis and the wave vector {\bf p} of quasiparticles.  
Namely, the coupling constant $\lambda_{{\bf q}}({\bf p})$ between the collective 
mode with wave vector {\bf q} and an electron with wave vector {\bf p} is given as 
\begin{equation}
\lambda_{{\rm q}}({\bf p})=\lambda_{\parallel}
({\hat p}_{x}Q_{xz,{\bf q}}+{\hat p}_{y}Q_{yz,{\bf q}})
+{\rm i}\lambda_{\perp}q_{z}{\hat p}_{z}
(Q_{xz,{\bf q}}+Q_{yz,{\bf q}}),
\label{coupling1}
\end{equation}
where ${\hat {\bf p}}\equiv {\bf p}/|{\bf p}|$ and $\lambda_{\parallel}$ and 
$\lambda_{\perp}$ are coupling constants that are essentially independent of 
{\bf q}.  

This reflects the fact that long-wavelength fluctuations of quadrupolar moment 
between two $\Gamma_{5}$ states around the equilibrium AF-Q order effectively 
scatter electrons travelling in the basal plane. On the other hand, 
these fluctuations give essentially no influence on electrons travelling 
perpendicular to the basal plane leaving the wave number of electrons intact.  
These situations can be understood intuitively from 
a picture shown in Fig.\ 1.  The $q$-linear term in the second term of 
eq. (\ref{coupling1}) represents the fact that the electrons 
can only be scattered by the difference in the rotation angle (at different lattice 
sites) of the quadrupolar moments from the equilibrium direction.  
The absence of the $q$-linear term in the first term of eq. (\ref{coupling1}) reflects 
the fact that even a uniform rotation of quadrupolar moment in the $ab$-plane 
can scatter the electrons travelling in the plane. 

\begin{figure}[h]
\begin{center}
\rotatebox{0}{\includegraphics[width=0.6\linewidth]{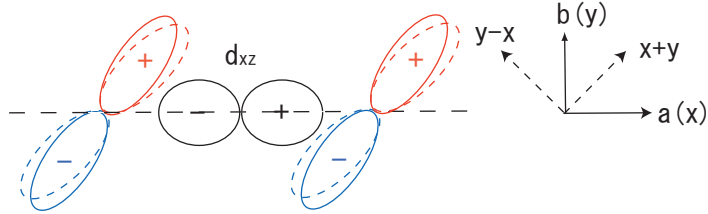}}
\caption{
Schematic view of how the coupling between quadrupole fluctuations of 
$Q_{z(x+y)}$ state and quasiparticles arises through modulation of the 
transfer integral via the $4d_{xz}$ orbital 
of Ru's located above or below the $ab$-plane including U ions.  
}
\label{Fig:2}
\end{center}
\end{figure}

Physical and microscopic origins of the couplings are primarily the modulation 
of the transfer integral of quasiparticles by the fluctuations of quadrupole in the 
$ab$-plane.  Figure \ref{Fig:2} shows how the transfer integral between adjacent 
U-sites along the $a$-axis among which Ru's are located above or below the $ab$-plane 
including U's. It is easily understood that the transfer of quasiparticles 
(consisting mostly of $f$ electrons) via the $d_{xz}$ orbital of Ru decreases if the 
$Q_{xz}$ component of the $f^{1}$ configuration of electrons at U ions increases from 
that of the original orbital shown by dashed lines.  This implies that 
the coupling arises between the quadrupole fluctuation $Q_{xz,{\bf q}}$, of wave vector 
${\bf q}$, and the electron transfer $c^{\dagger}_{{\bf p}+{\bf q}}c_{\bf p}$.  
This coupling works even for the uniform variation in $Q_{xz}$ component  at 
all lattice points.  On the other hand, the transfer integral 
along the $c$-axis between two adjacent $ab$-planes with the same distribution 
of U ions (which is not shown in the figure) does not change at all (in the 
linear approximation of changes in $Q_{xz}$ component) if the 
changes in $Q_{xz}$ component are the same at all lattice points. 
The coupling arises only from the difference in $Q_{xz}$ component change 
between the adjacent $ab$-planes.  Therefore, the coupling is proportional to 
$q_{z}$ in the long wavelength limit.  These facts are represented by the 
electron-quadrupole interaction Hamiltonian $H_{\rm e-q}$ as
\begin{eqnarray}
& &H_{\rm e-q}=\sum_{{\ell},z}\sum_{(i,j),(x,y)}
t^{\prime}_{ab}\left(Q_{xz}c^{\dagger}_{ij\ell}c_{i+1,j\ell}
+Q_{yz}c^{\dagger}_{ij\ell}c_{i,j+1,\ell}+{\rm h.c.}\right)
\nonumber
\\
& &\qquad\qquad
+\sum_{(i,j),(x,y)}\sum_{{\ell},z}
t^{\prime}_{c}\bigg[(Q_{xz,\ell}-Q_{xz,\ell+1})c^{\dagger}_{ij\ell}c_{ij,\ell+1}
\nonumber
\\
& &\qquad\qquad\qquad\qquad\qquad\qquad
+(Q_{yz,\ell}-Q_{yz,\ell+1})
c^{\dagger}_{ij\ell}c_{ij,\ell+1}+{\rm h.c.}\bigg],
\label{E-Q:2}
\end{eqnarray}
where $i$, $j$, and $\ell$ denote the lattice coordinates of $a$-, $b$-, and 
$c$-directions, respectively, and $t^{\prime}_{ab}$ and $t^{\prime}_c$ are constants 
parameterizing the changes of transfer integral in the $ab$-plane and the $c$-direction, 
respectively.  In eq. (\ref{E-Q:2}), only the hopping terms along the $a$-, $b$-, and 
$c$-directions are retained.  

%(consisting almost of $f$ electron)

\begin{figure}[h]
\begin{center}
\rotatebox{0}{\includegraphics[width=0.4\linewidth]{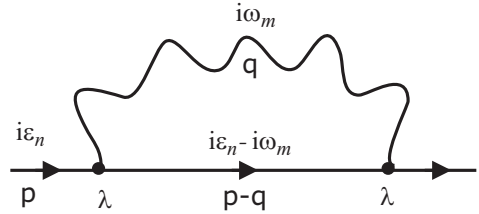}}
\caption{
Feynman diagram for the self-energy due to one-fluctuation mode exchange 
process.  
%$\lambda$ denotes the coupling between the AF-Q collective mode 
%and the quasiparticles.  
}
\label{fig:feyn}
\end{center}
\end{figure}

The self-energy due to one collective mode exchange process (given by the Feynman 
diagram shown in Fig.\ \ref{fig:feyn}) is 
\begin{equation}
\Sigma({\bf p},{\rm i}\epsilon_{n})=T\sum_{\omega_{m}}\sum_{\bf q}
|\lambda_{\bf q}|^{2}{1\over{\rm i}\epsilon_{n}-{\rm i}\omega_{m}
-\xi_{{\bf p}-{\bf q}}}\,{B\over \omega_{m}^{2}+\omega_{q}^{2}}.
\label{selfenergy1}
\end{equation}
After analytic continuation, ${\rm i}\epsilon_{n}\to \epsilon+{\rm i}\delta$, the 
imaginary part of the retardation self-energy, ${\rm Im}\Sigma^{\rm R}({\bf p},\epsilon)$, 
is calculated to give 
\begin{eqnarray}
& &{\rm Im}\Sigma^{\rm R}({\bf p},\epsilon)=-{\pi B\over 2}\sum_{\bf q}
{|\lambda_{{\bf q}}|^{2}\over 2\omega_{q}}
\Biggl[\left({\rm coth}{\omega_{q}\over 2T}-{\rm tanh}{\epsilon-\omega_{q}\over 2T}\right)
\delta(\epsilon-\xi_{{\bf p}-{\bf q}}-\omega_{q})
\nonumber
\\
& &
\qquad\qquad
+\left(\coth{\omega_{q}\over 2T}+\tanh{\epsilon+\omega_{q}\over 2T}\right)
\delta(\epsilon-\xi_{{\bf p}-{\bf q}}+\omega_{q})\Biggr].
\label{selfenergy2}
\end{eqnarray}
On the Fermi surface, i.e., ${\bf p}={\bf p}_{\rm F}$, and in the limit of 
$T\to 0$, eq. (\ref{selfenergy2}) is reduced to 
\begin{equation}
{\rm Im}\Sigma^{\rm R}({\bf p}_{\rm F},\epsilon)=
-{\pi B\over 2}\int_{0}^{|\epsilon|/v_{\rm F}}{{\rm d}q\over 4\pi^{2}}
{|\lambda_{{\bf q}}|^{2}q\over \omega_{q}v_{\rm F}}
\label{selfenergy3}
\end{equation}
Assuming the dispersion of the collective mode as $\omega_{q}\simeq sq$, 
eq. (\ref{selfenergy3}) is reduced to a rather simple form.  \\
For ${\bf p}\perp c$,
\begin{equation}
{\rm Im}\Sigma^{\rm R}({\bf p}_{\rm F},\epsilon)=
-{B\lambda_{\parallel}^{2}\over 8\pi^{2}}{1\over sv_{\rm F}}{|\epsilon|/v_{\rm F}},
\label{selfenergy4}
\end{equation}
and for ${\bf p}\parallel c$,
\begin{equation}
{\rm Im}\Sigma^{\rm R}({\bf p}_{\rm F},\epsilon)=
-{B\lambda_{\perp}^{2}\over 24\pi^{2}}{1\over sv_{\rm F}}
({|\epsilon|/v_{\rm F}})^{3}.
\label{selfenergy5}
\end{equation}
Since the scattering by the AF-Q collective mode of wave vector 
(${\bf Q}_{0}+{\bf q}$) couples with the Umklapp process, 
the temperature dependence of resistivity is given by 
${\rm Im}\Sigma^{\rm R}({\bf p}_{\rm F},\epsilon)$, where $|\epsilon|$ is replaced by 
$T$.  Namely, $\rho_{c}\propto T^{3}$ and $\rho_{a}\propto T$.  In the former case,  
the $T^{3}$ dependence is buried in $T^{2}$ term of the usual Fermi liquid behavior and 
will not be observed. 

Thus, the present note shows that the deviation from the $T^{2}$ term reported for 
$J\parallel a$ is a direct consequence of the nondipolar parameter of the hidden 
ordered phase.  The recent experimental work, which is also given by the current axis, 
also resolves that in many experiments, ithe $T^{2}$  behavior is not 
verified even t the temperature region near the superconducting transition 
temperature $T_{\rm SC}$\cite{Hassinger}.  
In agreement with our proposal, under pressures $P$ higher than $P_{x}$, where the 
ground state  switches from HO to AF, a good $T^{2}$ law is absent whatever the current 
axis ($a$ or $c$).  The difference from the case of AF order is that the coupling 
with quasiparticles is independent of the wave vector of the quasiparticles in the 
AF case, while it is highly anisotropic, as shown in eq.( \ref{coupling1}) in the present 
case of AF-Q order.  

In conclusion, we have succeeded in explaining, on an assumption that the HO 
is an AF-Q order with $\Gamma_{5}$ ($Q_{z(x+y)}$ and 
$Q_{z(x-y)}$) symmetry, the anisotropy in the temperature dependence of resistivity 
in the HO phase of URu$_2$Si$_2$.

\section*{Acknowledgments}
%The author acknowledges J. Flouquet for taking my attention to the present 
%problem.  
We are grateful to H. Harima for informative conversations on a possible 
hidden ordered state supporting our model. 
This work was supported by a Grant-in-Aid for Specially 
Promoted Research (20001004) and a Grant-in Aid 
for Scientific Research on Innovative Areas ``Heavy Electrons" (No.20102008) 
from the Ministry of Education, Culture, Sports, Science and Technology.  
One of us (J.F.) was supported by the Global COE program from the Japan Society 
for the Promotion of Science for staying at the Division of Materials Physics, 
Department of Materials Engineering Science, School of Engineering Science, 
Osaka University, where this work was initiated.

\end{document}